\documentclass[12pt,epsfig]{article}

\usepackage{graphicx}
\usepackage{subfig}
\usepackage{amsfonts}
\usepackage{amssymb}
\usepackage{amsmath}
\usepackage{placeins}
\usepackage{bbold}
\usepackage{hyperref}
\usepackage{xcolor}

\newskip\humongous \humongous=0pt plus 1000pt minus 1000pt
  \newif\ifdtup

\def\frac#1#2{ {{#1} \over {#2} }}

\def\beq{\begin{equation}}
\def\eeq{\end{equation}}

\newcommand{\beqa}{\begin{eqnarray}}
\newcommand{\eeqa}{\end{eqnarray}}
\newcommand{\ba}{\begin{array}}
\newcommand{\ea}{\end{array}}
\newcommand{\bmat}{\begin{pmatrix}}
\newcommand{\emat}{\end{pmatrix}}
\newcommand{\bcas}{\begin{cases}}
\newcommand{\ecas}{\end{cases}}

\begin{document}

\title
{Settling an old story: solution of the Thirring model in thimble regularization}

\author
{F.~Di~Renzo and K.~Zambello\\
\small{Dipartimento di Scienze Matematiche, Fisiche e Informatiche, Universit\`a di Parma} \\
\small{and INFN, Gruppo Collegato di Parma} \\
\small{I-43100 Parma, Italy}\\
}

\maketitle

\begin{abstract}
Thimble regularisation of lattice field theories has been proposed as
a solution to the infamous sign problem. It is conceptually very clean
and powerful, but it is in practice limited by a potentially very
serious issue: in general many thimbles can contribute to the
computation of the functional integrals. Semiclassical arguments would
suggest that the fundamental thimble could be sufficient to get the
correct answer, but this hypothesis has been proven not to hold true
in general. A first example of this failure has been put forward in
the context of the Thirring model: the dominant thimble approximation
is valid only in given regions of the parameter space of the
theory. Since then a complete solution of this (simple) model in
thimble regularisation has been missing. In this paper we show that
a full solution (taking the continuum limit) is indeed possible. It
is possible thanks to a method we recently proposed which de facto
evades the need to simulate on many thimbles.
\end{abstract}

\section{Introduction}
\label{sec:Introduction}
The lattice regularization of a Quantum Field Theory (QFT) enables
access to non-perturbati-ve aspects of the theory
by numerical simulations. On the lattice physical quantities are
defined as high-dimensional integrals which can be
computed via importance sampling by interpreting the factor $e^{-S}$ that appears in the integrands as a probability distribution.
This technique has proven to be very effective to understand, for instance, the phenomenon of confinement in QCD. 
Unfortunately there are very interesting theories whose action $S$ is
complex, so that $e^{-S}$ is not a well-defined
probability distribution. This is the so-called sign problem and it affects for instance QCD at finite density. For this very
reason the phase diagram of QCD remains largely unexplored by lattice simulations.

Not so long ago, following earlier work by Witten \cite{Witten1,Witten2}, it was proposed that one could evade the sign problem 
by regularizing a QFT on Lefschetz thimbles \cite{Aurora,Kikukawa}. The idea is simple. After complexifying the degrees of freedom 
one can define a set of manifolds attached to the critical
(stationary) points of the theory. These manifolds are called Lefschetz thimbles.
The Lefschetz thimbles are a basis for the original integration circle in the homological sense, i.e. the original integrals can be
decomposed to a linear combination of integrals over the thimbles.
In other terms the thimble decomposition defines a convenient
deformation of the original domain of integration (as a result 
the value of the integral does not change). It is a convenient deformation because on the thimbles the imaginary part of the action stays constant. Thus every
contribution is free of the sign problem, apart from a residual phase coming from the orientation of the thimble in the embedding manifold.
This gives rise to a residual sign problem, which is in practice 
found to be a (quite) milder one. Also, the coefficients
appearing in the decomposition (the so-called intersection numbers) are integers and they can be zero,
meaning that not all the thimbles do necessarily contribute.

Semiclassical arguments would suggest that the thimble attached to the
critical point having the minimum real part of the action (aka the
fundamental thimble) gives the
dominant contribution. This contribution is expected to be further enhanced in the continuum limit. Since the theory formulated
on a single thimble has the same symmetries and the same perturbative expansion as the original theory, it was conjectured
that the contribution coming from the fundamental thimble could be considered a good approximation for the complete theory.
This conjecture is known as the single-thimble dominance hypothesis. 
Soon it was discovered that this hypothesis holds true in some cases,
but does not hold true in general. Various counter-examples have been found, such as 
the ($0$+$1$)-dimensional QCD \cite{QCD01} and heavy-dense QCD \cite{HDqcd}, where the single-thimble approximation fails.
To our knowledge, the first counter-example that has been discovered is the one-dimensional Thirring model.
In refs. \cite{ThirringKiku,StudyThirringKiku,Alexandru:2015xva,Alexandru:2015sua} 
it was shown that taking into account the fundamental thimble alone is not enough to reproduce
the correct results, not even in the continuum limit. Indeed how to address the failure of the single-thimble approximation for this theory
has been an open question that we want to settle in this paper. There
is little doubt that a convenient thimble regularisation of the theory
will work, but this is expected to take into account contributions
from many thimbles and of course ``How many?'' is a key issue. We will
show that we can indeed compute the solution of the theory by thimble
regularisation (and the continuum limit can be taken). Quite interestingly, 
this will be most effectively achieved in a way that
to a large extent evades the subtleties of collecting contributions
from many thimbles.

The problem of collecting the contributions from more than one thimble is non-trivial. The contributions can be individually
estimated by importance sampling, but they must be properly weighted in the thimble decomposition. A universal and satisfactory
way to obtain the weight is to date still missing. This was a major motivation for
the exploration of alternative formulations somehow inspired by
thimbles, {\em e.g.} the holomorphic
flow \cite{Alexandru:2015sua,Fukuma:2017fjq} or various definitions of 
sign-optimised manifolds, possibly selected by deep-learning techniques 
\cite{PauloAndrei2,PauloAndrei3,Mori,ContourDefSUN}. Still, some proposals for
multi-thimbles simulations have been made in the literature \cite{QCD01,HDqcd,reweightHeidelberg,Bielefeld}.
In particular, in ref. \cite{HDqcd} we proposed a two-steps process in which one first computes the weights in the semiclassical approximation and then 
computes the relevant corrections. This worked fairly well for the simple version of heavy-dense QCD we had considered. 
Nonetheless it turned out that this method doesn't work that well for the Thirring model.
Therefore we turned back to the method we had thought of in ref. \cite{QCD01}, where the weights
are obtained by using a few known observables as normalization points. We also applied the method we have proposed
in ref. \cite{DiRenzo:2020cgp} in which observables are reconstructed by merging via Pad\'e approximants different Taylor
series carried out around points where the single-thimble approximation is a good one. This last method proved
to be more effective and it allowed to repeat the simulations towards the continuum limit.

The paper is organized as follows. In section \ref{sec:ThimblesMC} we discuss the strategy we've put in place to perform
one-thimble simulations, in particular summarizing the algorithm we have adopted for the Monte Carlo integration.
In section \ref{sec:ThirringSingleThimble} we report the numerical results we have obtained
for the Thirring model from one-thimble simulations. Not surprisingly we observe the failure of the single-thimble approximations
as it was already reported in refs. \cite{ThirringKiku,StudyThirringKiku,Alexandru:2015xva,Alexandru:2015sua}.
In section \ref{sec:ThirringMultiThimble} we show
how we can address the issue by collecting the contributions from the sub-dominant thimbles. The relative weight between the dominant
and the sub-dominant thimbles is obtained by the method proposed in ref. \cite{QCD01}. Finally in section ref. \ref{sec:ThirringTaylor}
we apply the strategy outlined in ref. \cite{DiRenzo:2020cgp}. This
strategy allowed to carry out the study of the continuum limit.

\section{Monte Carlo integration on thimbles} 
\label{sec:ThimblesMC}
The thimble decomposition for a given observable $\langle O \rangle$ can be stated as
\begin{subequations}
\label{eq:allVEV}
\begin{align}
\left\langle O \right\rangle \, & = \, Z^{-1} \, \int dx \; e^{-S(x)} \, O(x) \label{eq:basicComputation} \\
       & = \, \frac{\sum_{\sigma} \; n_{\sigma} \,
  e^{-i\,S_I(p_{\sigma})} \, \int_{\mathcal{J}_\sigma} dz \;
  e^{-S_R(z)}\; O(z)\; e^{i\,\omega(z)}}{\sum_{\sigma} \; n_{\sigma} \, e^{-i\,S_I(p_{\sigma})}\, \int_{\mathcal{J}_\sigma} dz \;
  e^{-S_R(z)}\; e^{i\,\omega(z)}} \label{eq:ThimbleDecomposition} \\
       & = \, \frac{\sum_{\sigma} \; n_{\sigma} \,
  e^{-i\,S_I(p_{\sigma})} \, Z_{\sigma} \left\langle O\,e^{i\,\omega} \right\rangle_{\sigma}
         }{\sum_{\sigma} \; n_{\sigma} \, e^{-i\,S_I(p_{\sigma})}\, \,
         Z_{\sigma} \left\langle e^{i\,\omega} \right\rangle_{\sigma}} \label{eq:ThimbleDecompositionBis}
\end{align}
\end{subequations}
Eq. \ref{eq:basicComputation} defines the physical quantity we want to calculate, where
$x$ is a short-hand notation for the fields configurations and $S(x) = S_R(x) + i S_I(x)$
is the complex-valued action of the system.

In eq. \ref{eq:ThimbleDecomposition} the real degrees of freedom have been complexified, i.e. $x \mapsto z$.
The integrals at the numerator and at the denominator have been rewritten as linear combinations of integrals over the 
stable thimbles $\mathcal{J}_\sigma$ attached to the critical points
$p_\sigma$ (in this context we call critical points 
the stationary points of the complexified action, i.e. the points such that $\partial_z S (p_\sigma) = 0$).
The stable thimble attached to a given critical point $p_\sigma$ is the
union of the solutions of the steepest-ascent (SA) equations stemming
from the critical point,
\begin{equation}
\label{eq:saeqs}
\frac{d}{dt}z_i = \frac{\partial \bar{S}}{\partial \bar{z}_i} \mbox{ , \hspace{0.25cm}} z_i (-\infty) = p_{\sigma,i} \mbox{ . }
\end{equation}
Along each SA path the real part of the action is always increasing. On the other hand
the imaginary part of the action stays constant, therefore the expressions
$e^{-i\,S_I(p_{\sigma})}$ have been factorized in front of the integrals.
The $n_\sigma$ are integer coefficients 
known as intersection numbers. These count the number of intersections between the original integration contour and
the unstable thimbles $\mathcal{K_\sigma}$, which are defined as the unions of the solutions of the steepest-descent (SD)
equations leaving the critical point. Along each SD path the real part of the action is always decreasing.
In eq. \ref{eq:ThimbleDecomposition} a so-called residual phase $e^{i\,\omega(z)}$ appears in the integrals.
This phase comes from the orientation of the thimble in the embedding manifold and it introduces
a residual sign problem, though this is usually found to be a mild one.

In eq. \ref{eq:ThimbleDecompositionBis} the thimble decomposition has been reformulated in a way that is amenable to numerical simulations.
We have defined an expectation value on the thimble $\mathcal{J}_\sigma$,
\begin{equation}\langle f \rangle_\sigma = \frac{\int_{\mathcal{J}_\sigma} dz ~ f ~ e^{-S_R}}{\int_{\mathcal{J}_\sigma} dz ~ e^{-S_R}} =
\frac{\int_{\mathcal{J}_\sigma} dz ~ f ~ e^{-S_R} }{Z_\sigma} \mbox{ , }\label{eq:thimblecontribution}\end{equation}
that we plan to evaluate stochastically. Notice that in order to make use of the thimble decomposition, one has to carry out two tasks.
In addition to computing the thimble contributions $\langle f \rangle_\sigma$ one also has to calculate their associated weights $Z_\sigma$.
Now we summarize the procedure we have put in place to numerically compute $\langle f \rangle_\sigma$. This is all that is needed
in the single-thimble approximation, where one only takes into account the contribution coming from the fundamental thimble.
Later on we will also discuss the issue of computing the weights in a
generic thimble decomposition.

By solving the Takagi's problem for the Hessian of the action $H(S;z)$ 
$$ H(S;p_\sigma)  \, v^{(i)} = \lambda_i \, \bar{v}^{(i)} $$
one can find a set of Takagi vectors $v^{(i)}$ and
their associated Takagi values $\lambda_i$. The Takagi vectors are a basis for the tangent space at the critical point.
After having fixed a normalization $\mathcal{R}$, the direction
associated to a given (infinitesimal) displacement from the critical point (a sort of initial
condition for a given SA path) can be expressed as $\sum n_i v^{(i)}$,
with $\hat{n} \in \mathcal{S}^{n-1}_\mathcal{R}$, where $\mathcal{S}^{n-1}_\mathcal{R}$ is the $(n-1)$-dimensional hypersphere of radius $\mathcal{R}$.
A point on the thimble can be singled out by the initial direction of the flow $\hat{n}$ and by the integration time $t$,
$$z \in \mathcal{J_\sigma} \leftrightarrow (\hat{n}, t) \in \mathcal{S}^{n-1}_\mathcal{R} \times \mathbb{R} \mbox{ . }$$

Let's first consider the denominator of eq. \ref{eq:thimblecontribution}, which we somehow improperly refer to as a partition
function. We can rewrite it in terms of the integration measure $\mathcal{D}\hat{n} = \prod_k dn_k \delta (|\hat{n}|^2 - \mathcal{R}^2)$,
$$Z_\sigma = \int_{\mathcal{J}_\sigma} dz ~ e^{-S_R(z)}  = \int \mathcal{D}\hat{n} ~ \int dt ~ \Delta_{\hat{n}}^\sigma(\hat{n},t) ~ e^{-S_R(\hat{n},t)}
\equiv \int \mathcal{D}\hat{n} ~ Z_{\hat{n}}^\sigma \mbox{ , }$$
where $\Delta_{\hat{n}}^\sigma(\hat{n},t)$ is a leftover from the change of variables. An explicit expression
for $\Delta_{\hat{n}}^\sigma(\hat{n},t)$ was worked out in ref. \cite{thimbleCRM}.  One finds the following expression
for the \textit{partial} partition function, 
$$Z_{\hat{n}}^\sigma = 2 (\sum_i \lambda_i n_i^2 ) \int dt ~ e^{-S_{eff}(\hat{n},t)} \mbox{ . }$$
Notice that we have defined the effective action $S_{eff}(\hat{n},t) = S_R(\hat{n},t) - log|det~V(\hat{n},t)|$ from the real
part of the action $S_R(\hat{n},t)$ and the matrix $V(\hat{n},t)$ having as columns the basis vectors of the tangent space at the time $t$.
In order to compute the partial partition function one has to parallel transport the basis vectors along the flow starting
from the critical point, where a basis is known (the Takagi vectors). This requires integrating the parallel transport equations
\begin{equation}
\label{eq:pteqs}
\frac{dV_j^{(h)}}{dt} = \sum_i \overline{V}_i^{(h)} \overline{\partial^2_{z_i z_j} S} \mbox{ . }
\end{equation}
The \textit{partial} partition function can be interpreted as the contribution to the partition function coming from an entire SA
path.  Similarly for the numerator one finds
$$ \int_{\mathcal{J}_\sigma} dz ~ f ~ e^{-S_R(z)} = \int \mathcal{D}\hat{n} ~ f_{\hat{n}}$$
$$f_{\hat{n}} = 2 (\sum_i \lambda_i n_i^2 ) \int dt ~ f(\hat{n},t) ~ e^{-S_{eff}(\hat{n},t)} \mbox{ . }$$
All in all the expectation value $\langle f \rangle_\sigma$ becomes
$$ \langle f \rangle_\sigma = \frac{\int \mathcal{D}\hat{n} ~ f_{\hat{n}}}{Z_\sigma} = \int \mathcal{D}\hat{n} ~ \frac{Z_{\hat{n}}^\sigma }{Z_\sigma} ~ \frac{f_{\hat{n}}}{Z_{\hat{n}}^\sigma}$$
This expectation value can be computed by importance sampling, sampling entire SA paths $\propto \frac{Z_{\hat{n}}^\sigma }{Z_\sigma}$
and estimating the result from the sample mean $\frac{1}{N}
\sum_{\hat{n}} \frac{f_{\hat{n}}}{Z_{\hat{n}}}$. All in all, in the
following we have
to think of a SA (parametrised by a direction $\hat{n}$) as we think
of a configuration in a standard Monte Carlo. 
The algorithm proceeds in two steps.
\begin{enumerate}
\item First we make a Metropolis proposal starting from the previous
  {\em configuration} $\hat{n}$. The proposal is generated
      by making $N$ consecutive rotations in planes defined by random directions. That is we pick two random integers $i \neq j$ and
      we perform a rotation in the place singled out by the $i$-th and the $j$-th direction, i.e. we (propose an) update $(n_i, n_j) \mapsto (n_i', n_j')$
      subject to the normalization condition. Then we pick another random integer $k \neq j$ and we perform a rotation
      in the plane singled out by the $j$-th and the $k$-th direction. We iterate until we have made $N$ rotations. Each rotation is obtained by parametrizing
      $(n_i, n_j)$ as
      $$
\begin{cases}
n_i = \sqrt{C} ~ sin(\phi) \\
n_j = \sqrt{C} ~ cos(\phi) \\
\end{cases}
$$
and proposing
$$
\begin{cases}
n'_i = \sqrt{C} ~ sin(\phi + \phi_0) \\
n'_j = \sqrt{C} ~ cos(\phi + \phi_0) \mbox { , } \\
\end{cases}
$$
where $C = n_i^2 + n_j^2$ and $\phi_0$ has been uniformly extracted in $[- \alpha, \alpha ]$.

\item Once we have proposed a new {\em configuration} $\hat{n}'$, we perform a Metropolis accept/reject test
by accepting the proposal with probability $P_{acc}(\hat{n}' \leftarrow \hat{n})
= min \left(1, \frac{Z_{\hat{n}'}^\sigma}{Z_{\hat{n}}^\sigma} \right) $.
Since the Metropolis proposal is symmetric, this is enough to satisfy the detailed balance principle and
the Markov chain will converge to the targeted probability distribution $\frac{Z_{\hat{n}}^\sigma }{Z_\sigma}$.

\end{enumerate}
The key ingredients that we need to compute are the partial partition function $Z_{\hat{n}}$
and the (contribution from the current SA path to the) observable $f_{\hat{n}}$.
These are time integrals whose computation requires to integrate the differential equations for both the fields
and the basis vectors given in eq. \ref{eq:saeqs} and eq. \ref{eq:pteqs}
starting from an initial condition close to the critical point,
$$\begin{cases}
z = z_\sigma + n_i e^{\lambda_i t} v^{(i)}\\
V^{(h)} = v^{(h)} e^{\lambda_h t} \mbox{ . }
\end{cases}
$$
Actually in our calculations we integrate in action instead of integrating in time. Since the real part of the action is monotonic in time,
one can make the change of variable
$$\frac{dS_R}{dt} = \frac{1}{2}\frac{d}{dt}(S + \overline{S}) = |\nabla S|^2 \mapsto dt = dS_R ~ |\nabla S|^{-2} = ds ~ |\nabla S|^{-2} \mbox{ . }$$
In the last step we have also defined a second change of variable $s = S_R - S_R(z_\sigma)$. In terms of $s$,
the differential equations for the fields and the basis become
$$\frac{dz_i}{ds} = |\nabla S|^{-2} ~ \frac{\partial \overline{S}}{\partial \overline{z}_i}$$
$$\frac{dV_j^{(h)}}{ds} = |\nabla S|^{-2} ~ \sum_i \overline{V}_i^{(h)} \overline{\partial^2_{z_i z_j} S} \mbox{ . }$$
The change of variable is also performed for the time integrals defining $Z_{\hat{n}}$ and $f_{\hat{n}}$, yielding
\begin{eqnarray}
\label{eq:eqinaction}
Z_{\hat{n}} = 2 \sum_i \lambda_i n_i^2 e^{-S_R(z_\sigma)} \int_0^\infty ds ~ |\nabla S|^{-2} ~ e^{-s + log|det~V(s)|} \\
f_{\hat{n}} = 2 \sum_i \lambda_i n_i^2 e^{-S_R(z_\sigma)} \int_0^\infty ds ~ f ~ |\nabla S|^{-2} ~ e^{-s +  log|det~V(s)|} \nonumber \mbox{ . }
\end{eqnarray}
The advantage is twofold. First, the integration in action has the effect of decreasing the number of iterations required to reach convergence.
Second, one can immediately recognize from eqs. \ref{eq:eqinaction} that the integrals can be calculated using
the Gauss-Laguerre quadrature, i.e.
$$\int_0^\infty f(x) e^{-x} dx = \sum w_i f(x_i) \mbox{ . }$$
where $\{x_i\}$ and $\{w_i\}$ are the quadrature points and their associated weights. The calculation of the determinant
(which is computationally quite expensive) is required only at the quadrature points.

\section{One-thimble simulations}
\label{sec:ThirringSingleThimble}
We now discuss the thimble regularization of the one-dimensional Thirring model.
As we have remarked in the introduction, historically this is one of the very first examples that have shown the
inadequacy of the single-thimble approximation (see refs \cite{ThirringKiku,StudyThirringKiku,Alexandru:2015xva,Alexandru:2015sua}).
The lattice action for this theory can be written down as
$$ S = \beta \sum_{n=1 \ldots L} (1 - cos(\phi_n)) - log~det~D \mbox{,}$$
where $det~D = \frac{1}{2^{L-1}} \left(cosh(L \hat{\mu} + i \sum_n \phi_n) + cosh(L~asinh(\hat{m})) \right)$ is the fermionic
determinant. The parameters $\hat{\mu} = \mu a$ and $\hat{m} = ma$ are respectively the chemical potential and the fermion mass in lattice
units, $\beta = (2g^2a)^{-1}$ is the inverse coupling constant and $\phi_n$ is a scalar field discretized on a one-dimensional lattice
of length $L$. The theory features a sign problem originating from the fermionic determinant, which is complex at finite $\hat{\mu}$.

An analytical solution is known for the partition function. This is given in term of the modified Bessel functions of the first kind $I_n(x)$,
%$$Z = \int \left( \prod_{i=1 \ldots L} d\phi_n \right) e^{- \beta \sum_{n=1 \ldots L} (1 - cos(\phi_n)) + log~detD}\mbox{,}$$
%$$detD = \frac{1}{2^{L-1}} \left(cosh(L \hat{\mu} + i \sum_n \phi_n) + cosh(L~asinh(\hat{m})) \right) \mbox{ . }$$

$$Z = \frac{1}{2^{L-1}} e^{-L\beta} \left[I_1(\beta)^L cosh(L\hat{\mu}) + I_0(\beta)^L cosh(L~asinh(\hat{m}))\right] \mbox{ . }$$

From the partition function one can derive closed-form expressions for the physical observables, such the
scalar condensate $\langle \bar{\chi}\chi \rangle$ and the fermion density $\langle n \rangle$,

$$\langle \bar{\chi}\chi \rangle = \frac{1}{L} \frac{\partial log~Z}{\partial \hat{m}}= \frac{1}{cosh(asinh(\hat{m}))} \frac{I_0(\beta)^L sinh(L~asinh(
\hat{m}))}{I_1(\beta)^L cosh(L \hat{\mu}) + I_0(\beta)^L cosh(L~asinh(\hat{m}))} \mbox{ . }$$

$$\langle n \rangle = \frac{1}{L} \frac{\partial log~Z}{\partial \hat{\mu}} = \frac{I_1(\beta)^L sinh(L \hat{\mu})}{I_1(\beta)^L cosh(L \hat{\mu}) + I_
0(\beta)^L cosh(L~asinh(\hat{m}))}$$

A solution by numerical methods, on the other hand, is difficult to obtain because of the sign problem.
In this paper we explore the feasibility of using the thimble regularization method to study the Thirring model.
The first step consists in complexifying the degrees of freedom. In this case, each real degree of freedom
$\phi_n$ is replaced by a complex degree of freedom $z_n$. The action is now given in terms of $z_n$,

$$ \beta \sum_{n=1 \ldots L} (1 - cos(z_n)) - log~detD \mbox{ , }$$

where $detD = \frac{1}{2^{L-1}} \left(cosh(L \hat{\mu} + i \sum_n z_n) + cosh(L~asinh(\hat{m})) \right)$.
The critical points of the theory are found by requiring a vanishing gradient,

\begin{equation}
\frac{\partial S}{\partial z_n} = \beta sin(z_n)  -  i~\frac{sinh(L \hat{\mu} + i \sum_i z_i)}{cosh(L \hat{\mu} + i \sum_i z_i) + cosh(L ~ asinh(\hat{m}))} = 0 \mbox{ . }
\label{eq:zerograd}
\end{equation}

%$$sin(z_n) = \frac{i}{\beta} \frac{sinh(L \hat{\mu} + i \sum_i z_i)}{cosh(L \hat{\mu} + i \sum_i z_i) + cosh(L ~ asinh(\hat{m}))} \mbox{ . }$$

From this condition one can see that $sin(z_n)$ takes for all $n=1
\ldots L$ always the same value, which we denote by
$sin(z)$\footnote{Here and in the following we adhere to the notation of \cite{ThirringKiku}.}
and which depends on $z_n$ only through the sum $\sum_n z_n$. Therefore the critical points are given by field configurations
where $z_i = z$ for any $i$ but a number $n_-$ of lattice points where $z_i$ can take the value $\pi - z$ (without changing $sin(z_n)$).
For a fixed $n_-$ the values admitted for $z$ are found by numerically solving

\begin{equation}
sin(z) = \frac{i}{\beta} \frac{sinh(L \hat{\mu} + i (L - 2n_-) z + i n_- \pi)}{cosh(L \hat{\mu} + i (L - 2n_-) z + i n_- \pi) + cosh(L~asinh(\hat{m}))} \mbox { . }
\label{eq:zerograd2}
\end{equation}

%$$ =  \frac{i}{\beta} \frac{sinh(L \hat{\mu} + i (L - 2n_-) z)}{cosh(L \hat{\mu} + i (L - 2n_-) z ) + (-1)^{n_-} cosh(L ~ asinh(\hat{m}))} \mbox{ . }$$

\begin{figure}[htb]
        \centering
        \includegraphics[scale=0.4]{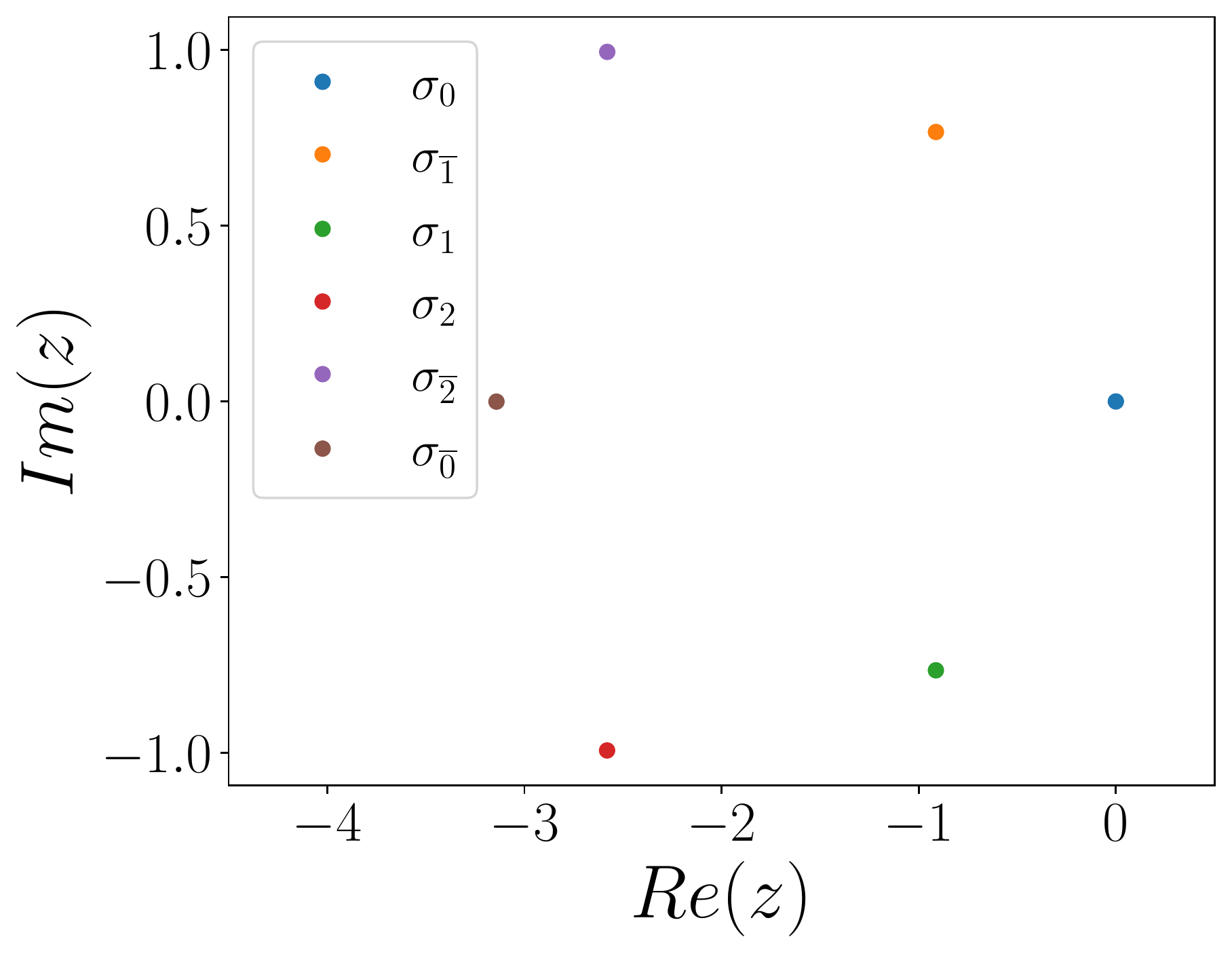}
        \includegraphics[scale=0.4]{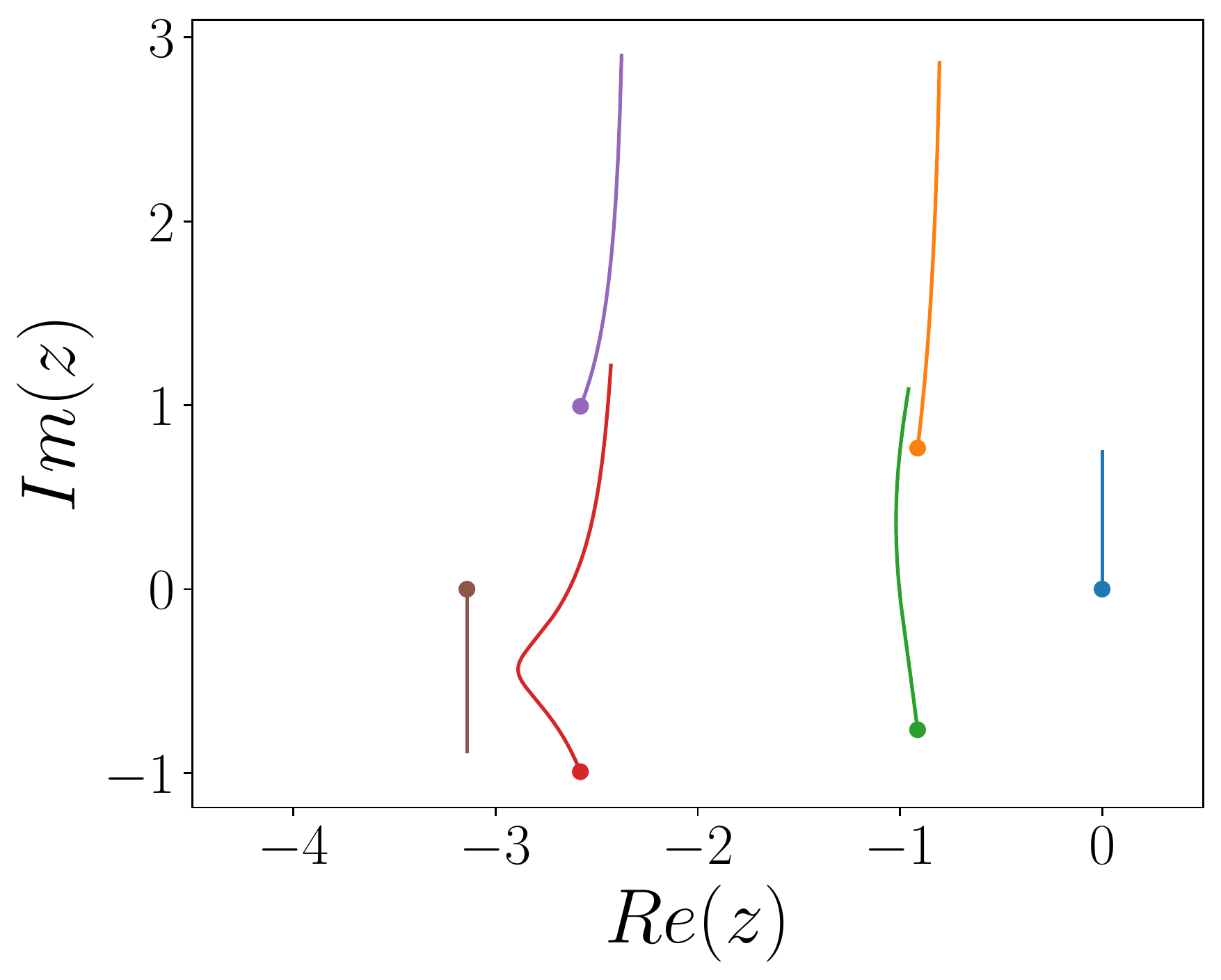} \\ \vspace{0.5cm}
        \includegraphics[scale=0.4]{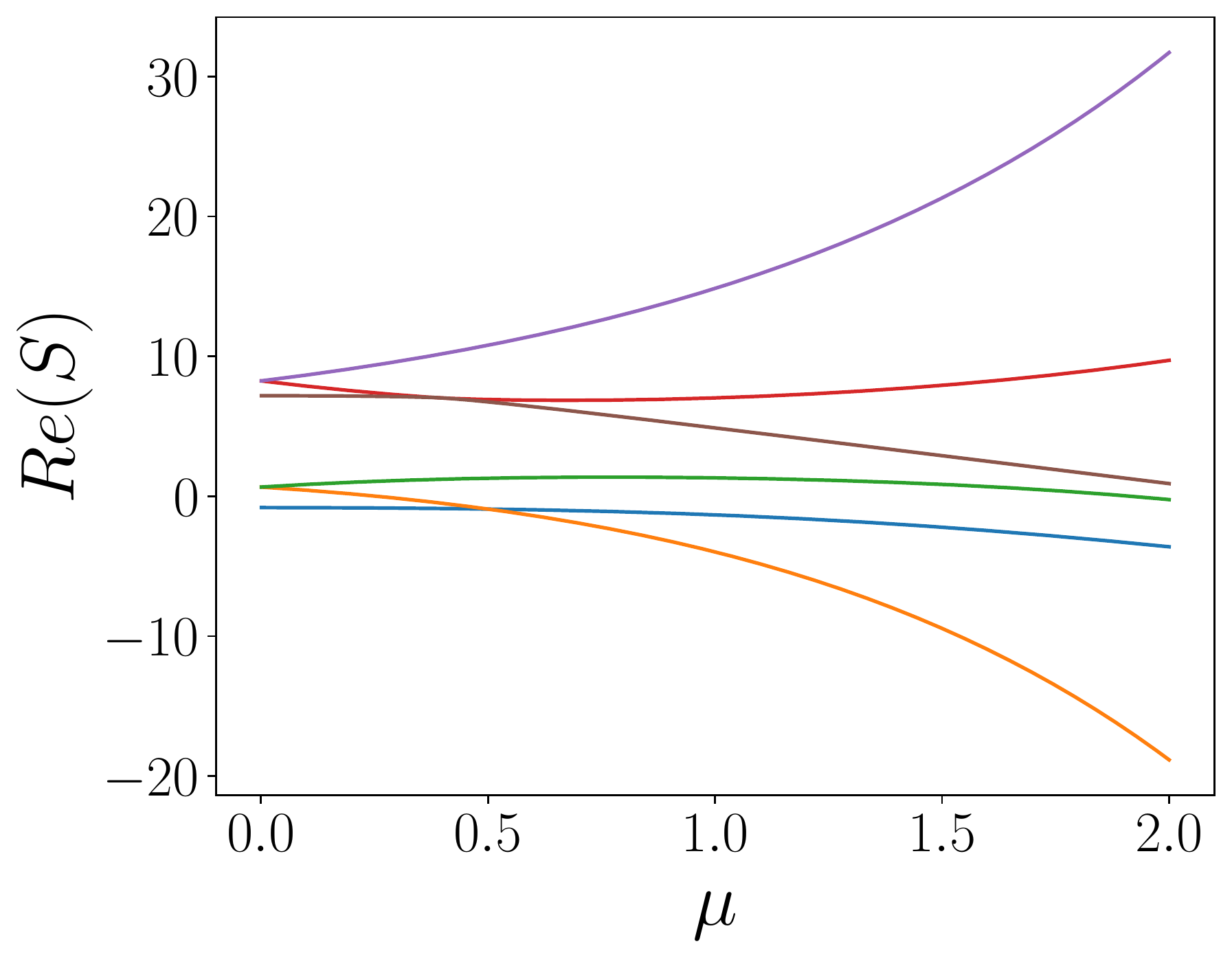}
        \includegraphics[scale=0.4]{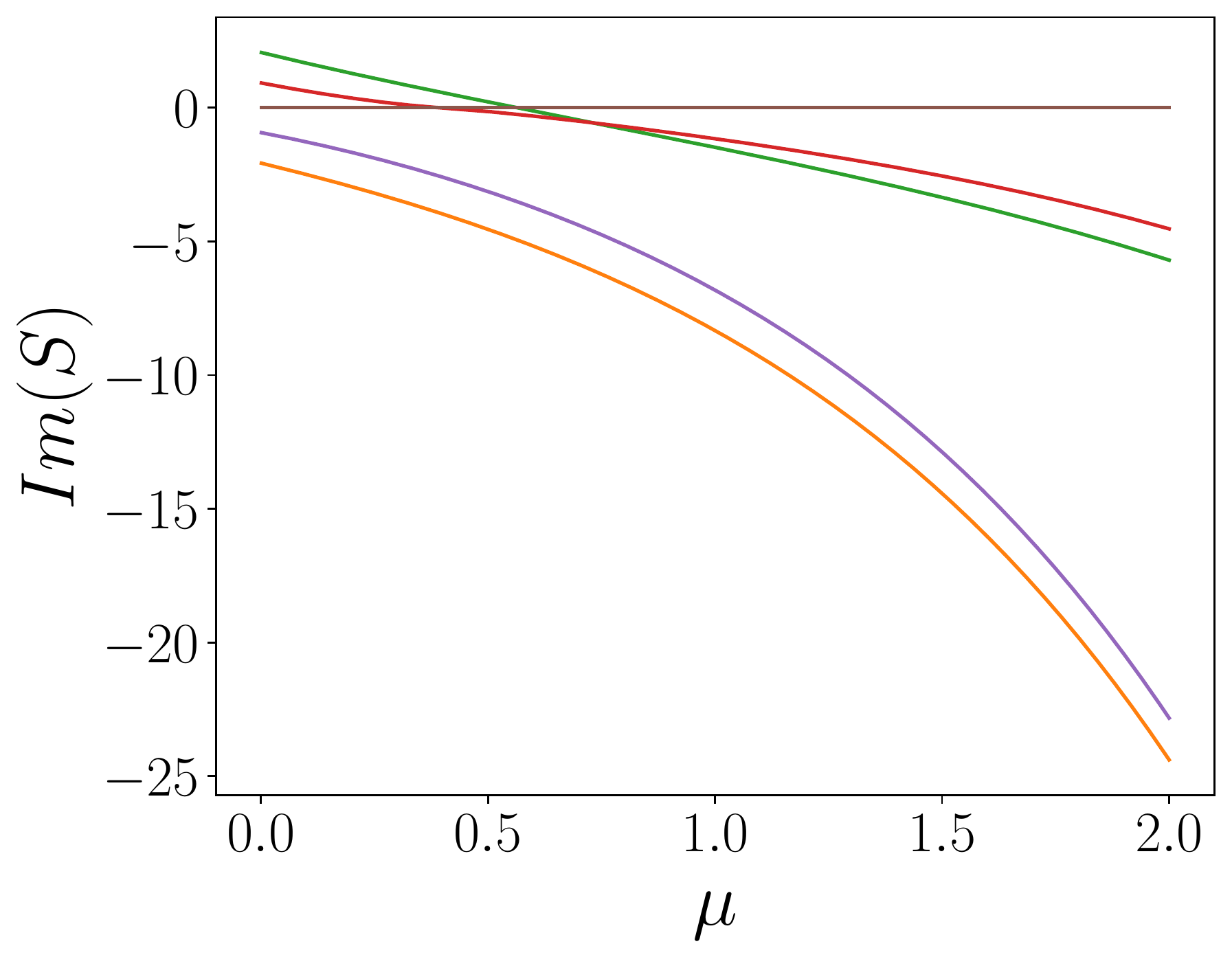}
        \caption{\label{fig:critp} Critical points for $L=4$, $\beta=1$ and $ma = 1$: solutions for $\hat{\mu} = 0$ (top left), solutions for $\hat{\mu} \in [0.0, ~2.0]$ (top right), real part of the action as a function of $\hat{\mu}$ (bottom left) and imaginary part of the action as a function of $\hat{\mu}$ (bottom right).}
\end{figure}

This equation follows directly from eq. \ref{eq:zerograd}. For instance let's consider the critical points 
in the $n_- = 0$ sector, which are expected to give the leading contributions \cite{StudyThirringKiku}.
The solutions corresponding to the parameters $L=4$, $\beta=1$ and $ma=1$ are graphically shown in fig. \ref{fig:critp}.
The top left figure shows the solutions for $z$ of eq. \ref{eq:zerograd2}, while
the top right figure shows how they move when we add a finite chemical potential.

\FloatBarrier
\FloatBarrier

Notice that only the left complex half-plane is shown. Since eq. \ref{eq:zerograd2} is invariant under $z \mapsto - \bar{z}$,
if $z$ defines a critical point so does $-\bar{z}$. Therefore each critical point $\sigma_i$ has a counterpart $\sigma_{-i}$
in the right complex half-plane. However these mirrored critical point do not give rise to independent contributions. Indeed 
the action displays a symmetry $S(-\bar{z}) = \overline{S(z)}$ which ensures that the thimble contributions from the member
of each pair $\sigma_{i}$, $\sigma_{-i}$ are conjugate contributions.

If we look at the bottom left picture of fig. \ref{fig:critp} we can see that the fundamental thimble, the one attached to the critical point
having the minimum (real part of the) action, is $\mathcal{J}_{\sigma_0}$. Actually from some chemical potential $\mu_0$ onwards
the critical point $\sigma_{\bar{1}}$ has an even lower (real part of the) action, but since this is also lower than
the minimum real part of the action on the original domain, the thimble attached
to this critical point cannot appear in the thimble decomposition.

\begin{figure}[h!]
        \centering
        \includegraphics[scale=0.4]{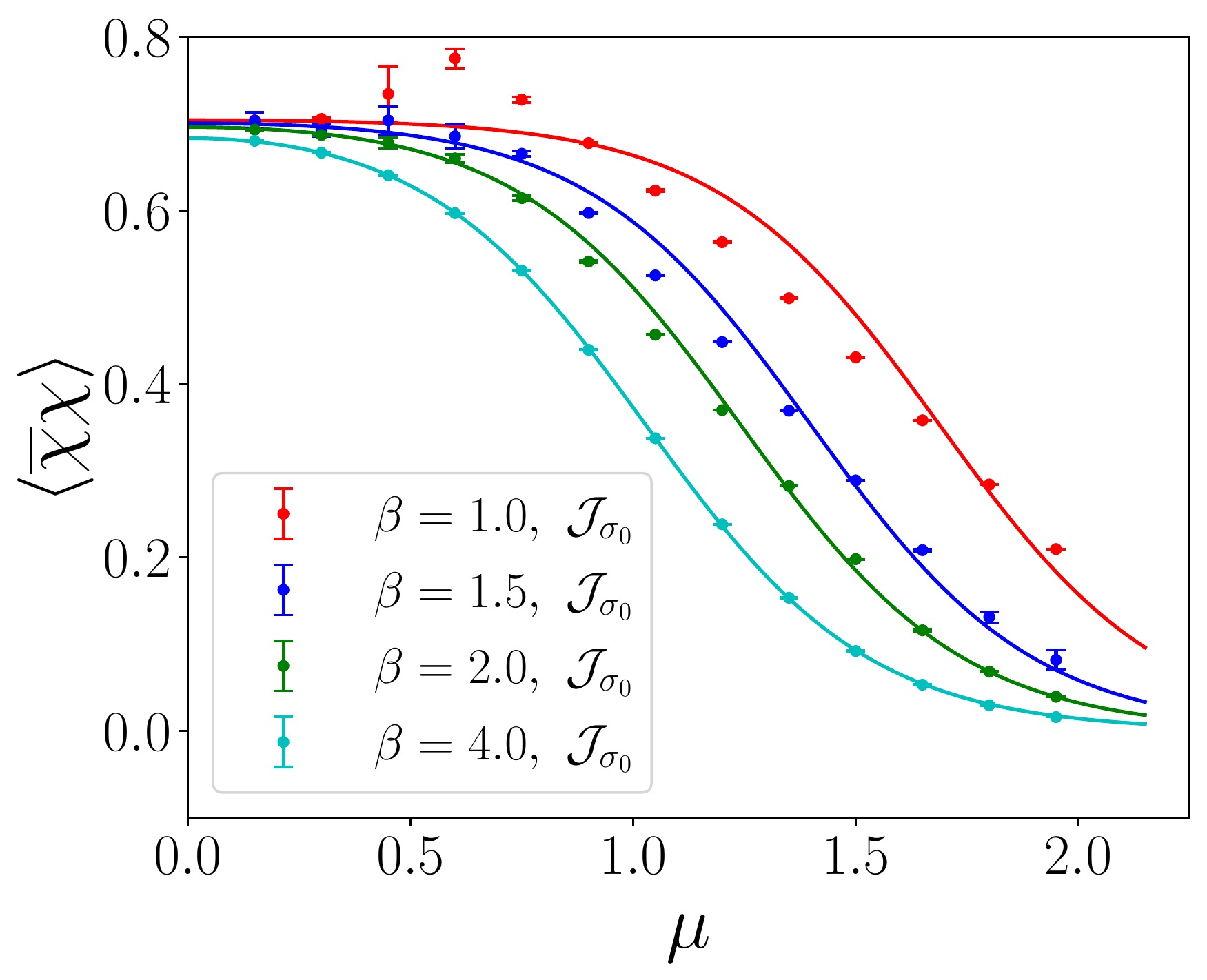}
        \caption{\label{fig:res1t} Scalar condensate for $L=4$, $\beta=1.0$, $1.5$, $2.0$, $4.0$ and $ma = 1$: results from one-thimble simulations on $\mathcal{J}_{\sigma_0}$.}
\end{figure}

In fig. \ref{fig:res1t} we show the numerical results obtained from numerical simulations performed on the fundamental thimble $\mathcal{J}_{\sigma_0}$
for different values of $\beta=1.0$, $1.5$, $2.0$, $4.0$. \footnote{Actually a small imaginary part has been added to $\beta$ in order to prevent a Stokes
phenomenon between $\mathcal{J}_{\sigma_0}$ and $\mathcal{J}_{\sigma_{\bar{0}}}$.}
At strong couplings (i.e. at low $\beta$) the single-thimble approximation yields the wrong results, at least for high chemical
potentials. This does not come as a surprise (it is exactly what other
authors found previously) and shows that the contribution from the sub-dominant thimbles cannot be neglected.

\FloatBarrier
%\begin{figure}[tbp]
\FloatBarrier

\section{Multi-thimble simulations}
\label{sec:ThirringMultiThimble}

The reason why the single-thimble approximation fails is that, when the chemical potential is increased starting from zero,
different Stokes phenomena take place and the thimble decomposition changes.
As a result thimbles other than the fundamental one need to be taken into account at high chemical potentials,
even though at zero chemical potential the contribution from the fundamental thimble $\mathcal{J}_{\sigma_0}$ is
the only non negligible one.

All in all the sub-dominant contribution that one has to take into account is the one from the thimble
$\mathcal{J}_{\sigma_1}$, which enters the decomposition at $\hat{\mu} \approx 0.56$. Indeed at this chemical
potential the imaginary parts of the action at $\sigma_0$ and $\sigma_1$ are the same, as shown in the
bottom right picture of fig. \ref{fig:critp}. This is a necessary condition for a Stokes phenomenon, that indeed
happens. For a very nice and thorough analysis of the Stokes phenomena and their consequences on the thimble decomposition, 
we refer the reader to ref. \cite{StudyThirringKiku}. 

In order to properly collect the contributions from the fundamental thimble $\mathcal{J}_{\sigma_0}$ and
the sub-dominant thimble $\mathcal{J}_{\sigma_1}$, we have to determine the relative weights of the two
contributions. The approach we took is the one we've followed for ($0$+$1$)-dimensional QCD. 
Since we are considering only two independent thimble contributions, the expectation value of a generic observable $O$
can be written as
$$\langle O \rangle = \frac{n_0 e^{-iS_I(z_0)} Z_0 \langle O e^{i \omega_0} \rangle_0
+ n_{12} e^{-iS_I(z_{12})} Z_{12} \langle O e^{i \omega_{12}} \rangle_{12}}
{n_0 e^{-iS_I(z_0)} Z_0 \langle O e^{i \omega_0} \rangle_0
+ n_{12} e^{-iS_I(z_{12})} Z_{12} \langle O e^{i \omega_{12}} \rangle_{12}} \mbox{ . }$$

Here we use the subscript $12$ to denote quantities that refer to the critical points $\sigma_{1}$ and $\sigma_{-1}$.
As we have already observed these critical points give rise to conjugate contributions (whose
sum is purely real). When $\mathcal{J}_{\sigma_1}$ enters the thimble decomposition,
so does $\mathcal{J}_{\sigma_{-1}}$, but we have only one independent contribution.
Now if we divide both the numerator and the denominator by $n_0 e^{-iS_I(z_0)} Z_0$ we obtain
$$\langle O \rangle = \frac{\langle O e^{i \omega_0} \rangle_0 + \alpha \langle O e^{i \omega_{12}} \rangle_{12}}
{\langle e^{i \omega_0} \rangle_0 + \alpha \langle e^{i \omega_{12}} \rangle_{12}} \mbox{ . }$$
where $\alpha = \frac{n_{12} e^{-iS_I(z_{12})} Z_{12}}{n_0 e^{-iS_I(z_0)} Z_0}$.
Since $\alpha$ only depends on the thimble structure of the theory, we can determine it by taking some observable
$\tilde{O}$ as a normalization point and then use such value to calculate any other observable.

\begin{figure}[h!]
        \centering
        \includegraphics[scale=0.4]{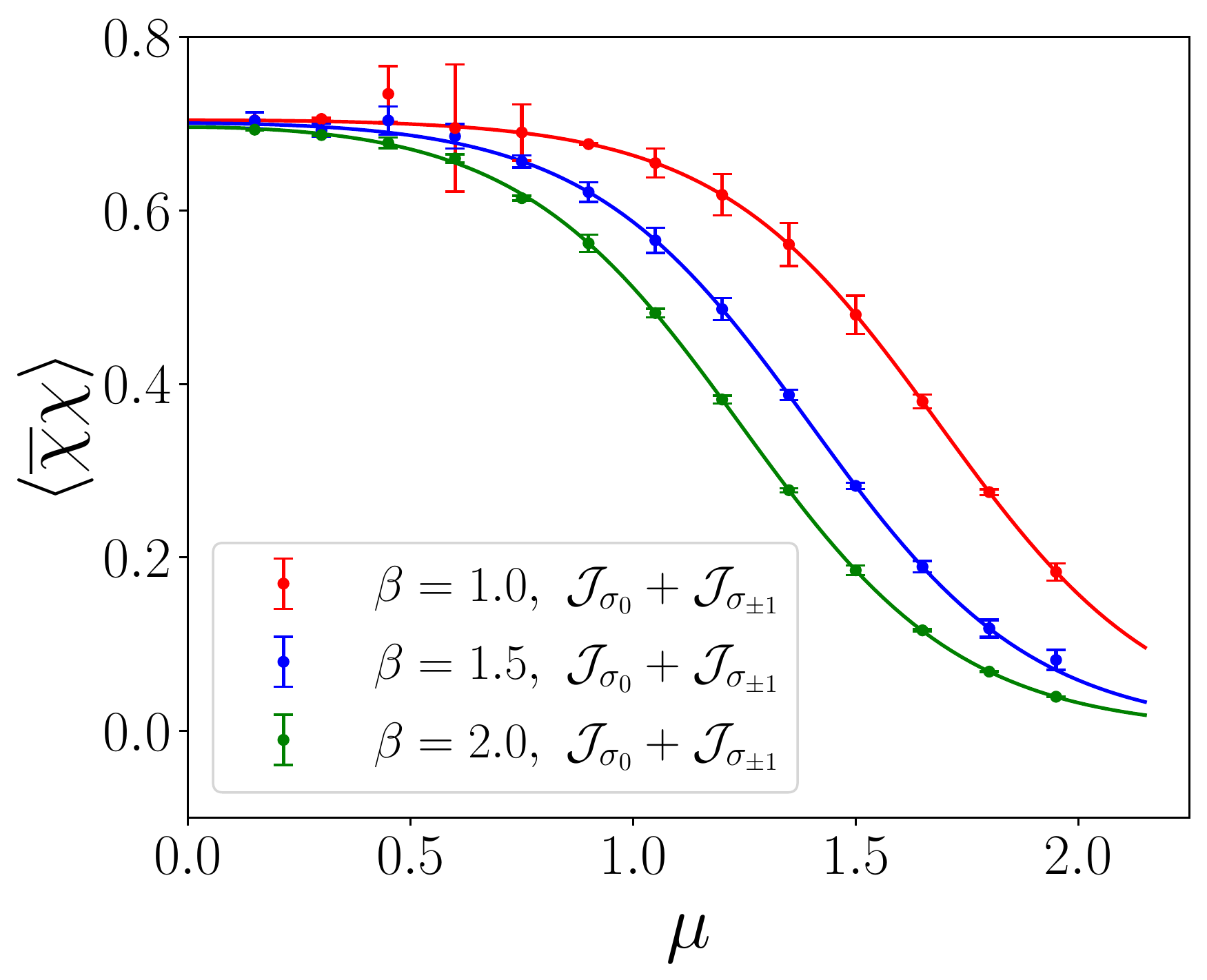}
        \caption{\label{fig:res2t} Scalar condensate for $L=4$, $\beta=1.0$, $1.5$, $2.0$ and $ma = 1$: results from multi-thimble simulations on 
$\mathcal{J}_{\sigma_0}$ and $\mathcal{J}_{\sigma_{\pm 1}}$.}
\end{figure}

\begin{figure}[h!]
        \centering
        \includegraphics[scale=0.4]{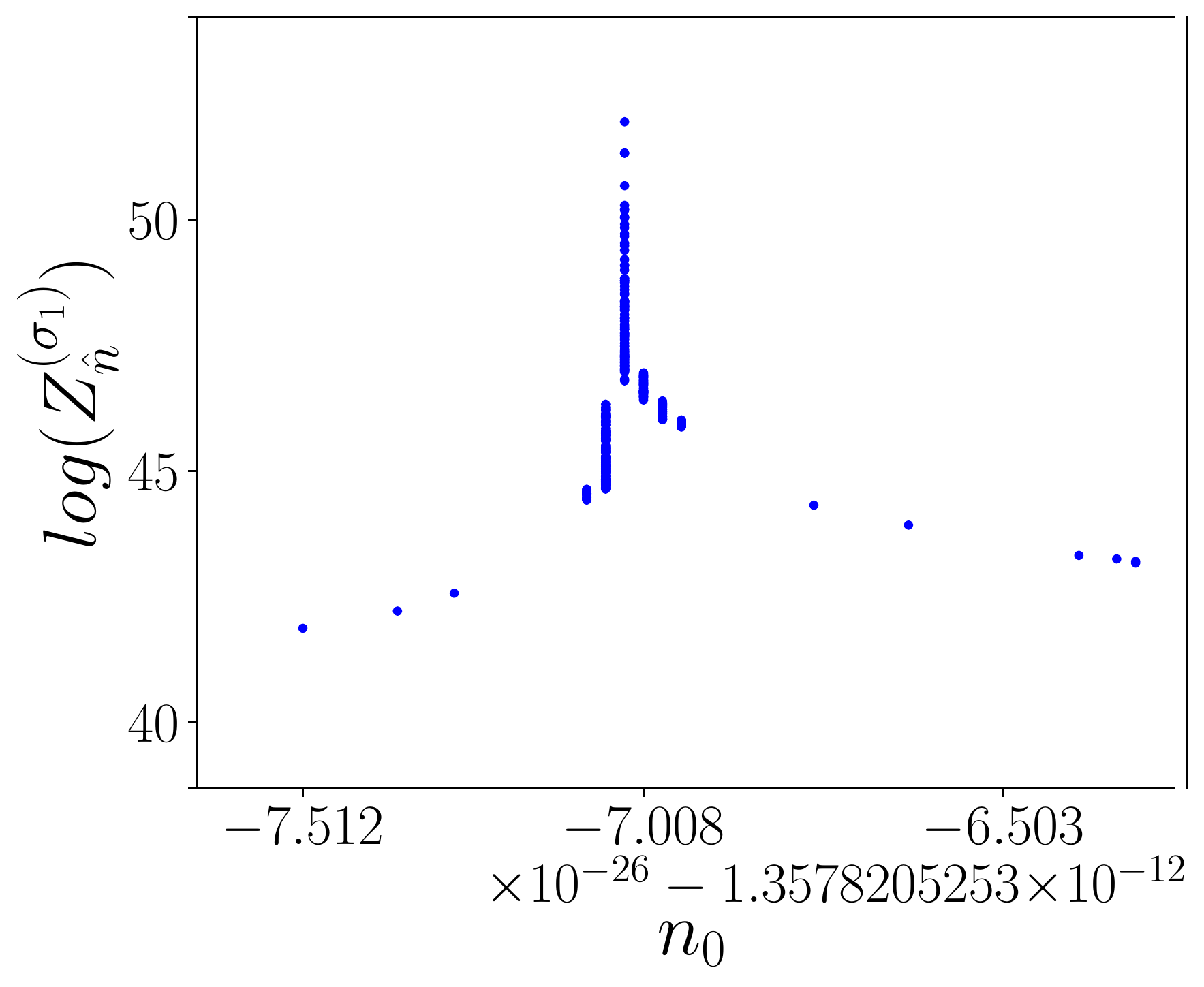}
        \caption{\label{fig:logZn} Logarithm of the partial partition function for the critical point $\sigma_1$. This is 
                                 plotted as a function of $n_0$, i.e. the initial direction
                                 of the SA path on the tangent space along the Takagi vector having the largest Takagi value. Parameters: $L=2$, $\beta = 1$ and $ma=1$.}
\end{figure}

For the Thirring model we fixed $\alpha$ from the (analytical solution of the) number density and we
used it to calculate the scalar condensate.
The numerical results are shown in fig. \ref{fig:res2t}. The results are now in agreement with the analytical solution:
the contribution from the sub-dominant thimble fully accounts for the discrepancies observed in the results
from one-thimble simulations (and that's why we could make a long
story short earlier, when we said that the sub-dominant contribution
that one has to take into account is the one from the thimble $\mathcal{J}_{\sigma_1}$).
Notice that the statistical errors are quite large for $\hat{\mu} \approx 0.6 \div 0.75$, this is in part due to cancellations in the
calculation of $\alpha$ and in part due to numerical difficulties in sampling the non-dominant thimble. 
The partial partition function (i.e. the probability distribution for importance sampling) shows sharp spikes
in some regions of $n_0$ (i.e. the initial direction of the SA path on the tangent space along the Takagi vector 
having the largest Takagi value). Within these regions the partial partition function varies by several
order of magnitude and this makes it difficult to keep a good acceptance ratio. Moreover for $\mathcal{J}_{\sigma_{1}}$ the regions are also so
thin that the regions of interest cannot represented in double precision and quadruple precision is needed in
the simulations, with a noticeable impact on the performance of the code. An example of this last issue
is shown in fig. \ref{fig:logZn}, where the logarithm of the partial partition function of $\mathcal{J}_{\sigma_{1}}$
is shown as a function of $n_0$ for a $L=2$ lattice.

In the next section we will study the Thirring model using a strategy
that will turn out to be much more effective, since {\em (a)} it does need
to take a known result as a normalisation point and {\em (b)} it does
not require to sample the sub-dominant thimble.

\FloatBarrier
%\begin{figure}[tbp]
\FloatBarrier

\FloatBarrier
%\begin{figure}[tbp]
\FloatBarrier

\section{Taylor expansions on the fundamental thimble}
\label{sec:ThirringTaylor}
In this section we follow the approach we proposed in
ref. \cite{DiRenzo:2020cgp} (the interested reader is referred to it
for further details on the method itself).
The main idea is that we can by-pass the need for multi-thimble simulations by calculating
multiple Taylor expansions around points where the single-thimble
approximation holds true.
As a consequence, the coefficients of such expansions can be computed by one-thimble simulations.
The method relies on the fact that while the thimble
decomposition can be (highly) discontinuous as we sample different
regions in the parameter space of the theory, physical observables (in
general) are not. Our strategy is to take advantage of the
single-thimble approximation holding true in given regions and bridge
these different (disjoint) regions by Taylor expansions. Actually it
turns out that the most efficient way to bridge these different regions
is by Pad\'e approximants. As a side-effect, having computed Pad\'e approximants, we are able
to probe the analytical structure of the observables, i.e. we can
locate their singularities. 
This makes the approach a very powerful tool for theoretical investigations,
with applications beyond thimble regularization itself.
In a more general framework, one could say that bridging different
regions by Pad\'e approximants can be an effective way of studying the
phase diagram of a theory for which direct computations can be
approached only in given regions of the parameter space.
Indeed such a strategy can be {\em e.g.} applied to lattice QCD at imaginary chemical
potential \cite{BielePR0,BielePR}.

From a numerical point of view the method of \cite{DiRenzo:2020cgp}
proved to be quite effective in our case. It allowed to study the Thirring model 
for (much) larger/finer lattices than the (modest) $L=4$ lattice we have considered in the previous sections.

\begin{figure}[h!]
        \centering
        \includegraphics[width=\textwidth, height=\textheight, keepaspectratio]{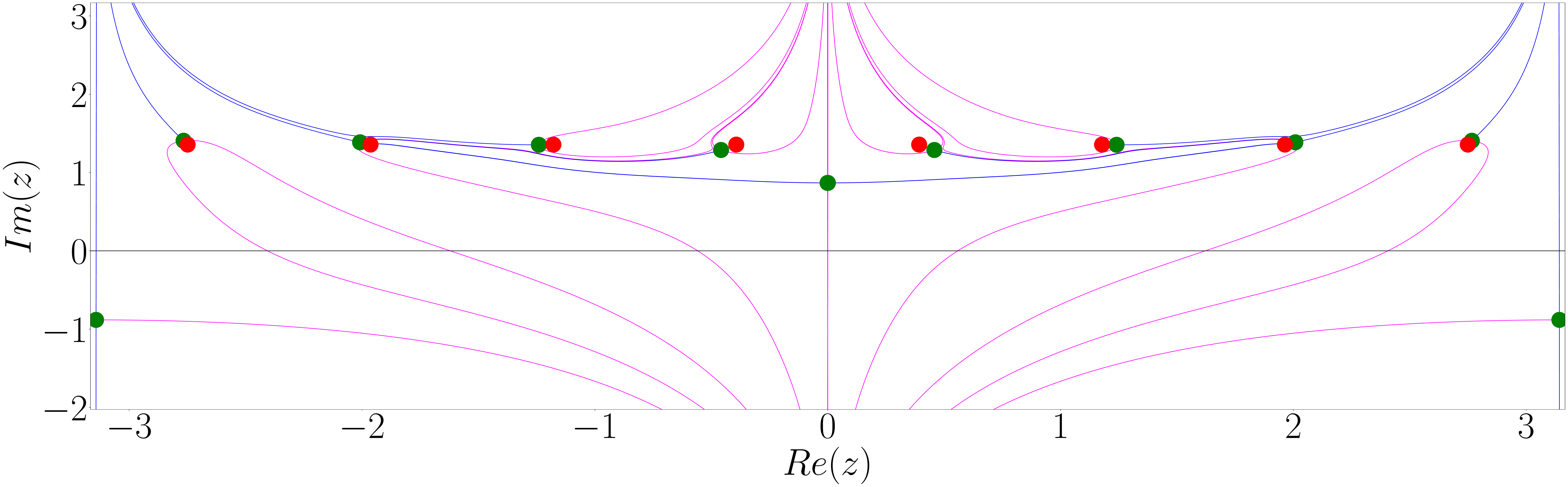}
        \includegraphics[width=\textwidth, height=\textheight, keepaspectratio]{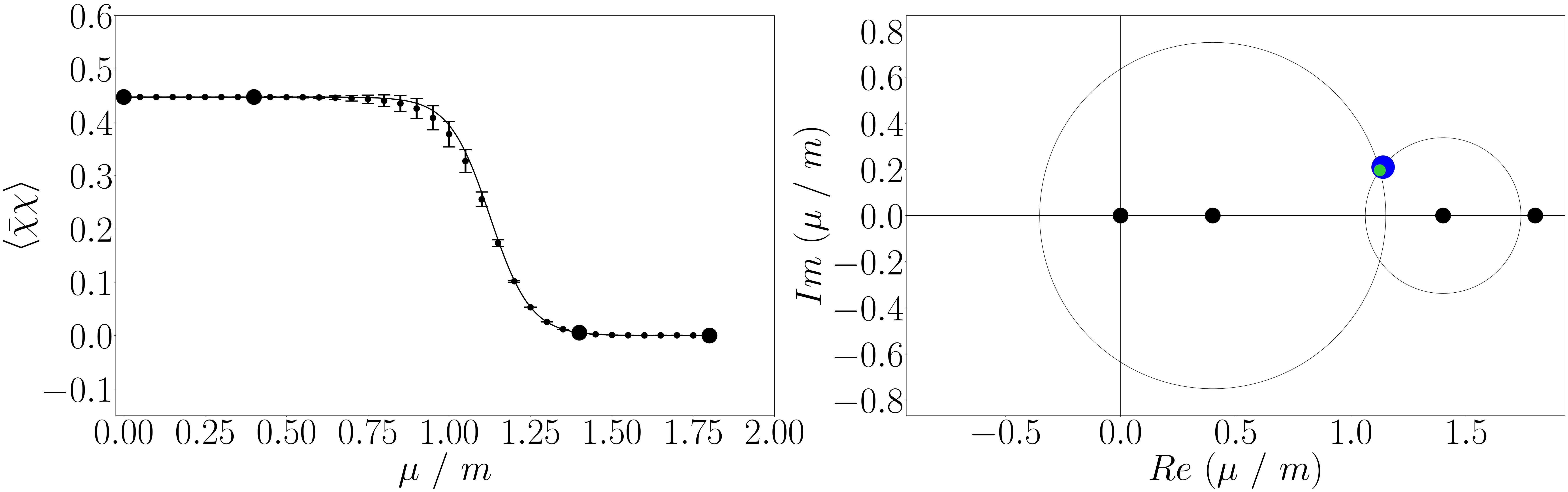}
        \caption{\label{fig:nr5} The top picture shows the thimble structure for $L=8$, $\beta = 1.0$ and $ma=2$ at $\frac{\mu}{m} = 1.4$.
                                 The critical points are represented by the green points, while the stable and unstable thimbles are
                                 displayed in blue and magenta. The critical point $\sigma_0$ is the one sitting at $Re(z) = 0$ and
                                 the critical point immediately on the left is $\sigma_1$. The unstable thimble attached to the former
                                 intersects the original domain of integration (the real axis), while the unstable thimble attached to
                                 the latter does not. The bottom left picture shows the results from Pad\'e as black error bars, which
                                 are in very good agreement with the expectations from the analytical solution (represented by the black solid line).
                                 The expansion points are also shown as black points. In the bottom right picture the expansion points
                                 are displayed on the complex $\frac{\mu}{m}$ plane. On top of these are also shown the singularity of the
                                 condensate (the green point) and the stable pole of the Pad\'e approximant (the blue point).}
\end{figure}

Let's start by studying the theory using $L=8$, $\beta = 1$ and $ma =
2$ as parameters. We need a few expansion points where
the Taylor expansion can be calculated by one-thimble simulations:

\begin{enumerate}
\item As a first point we selected $\frac{\mu}{m} = 0.4$. This choice
can be understood from a simple argument. The range of $S_I$ on the real 
domain of integration is limited. As a result, from explicit computation of
$S_I^{\sigma}(\frac{\mu}{m})$ we can conclude that below a given value
$\frac{\mu_0}{m}$ of the chemical potential there are only two unstable thimbles
that can intersect the real domain of integration. These are the one attached
to the fundamental critical point $\sigma_0$ and the one attached to
the critical point $\sigma_{\bar{0}}$. But the contribution
from $\mathcal{J}_{\sigma_{\bar{0}}}$ can be neglected, as $S_R(\sigma_{\bar{0}}) \gg S_R(\sigma_{0})$.

\item As a second point we selected $\frac{\mu}{m} = 1.4$. For this
value of the chemical potential, all but three the critical points
other than the fundamental one have $S_R(\sigma) \gg S_R(\sigma_{0})$. 
Hence we can neglect them. We denote the three remaining critical points 
by $\sigma_1$, $\sigma_{\bar{1}}$ and $\sigma_{\bar{2}}$. Two of them (namely $\sigma_{\bar{1}}$ and $\sigma_{\bar{2}}$)
have a real part of the action $S_R$ less than the minimum $S_R^{min}$ of the real action
on the original domain of integration. Hence their unstable thimble cannot intersect the original domain of integration.
As for $\sigma_1$ one can explicitly check that the attached unstable thimble does not intersect the original
domain of integration. See the top picture of fig. \ref{fig:nr5}. The critical point $\sigma_0$ is represented by
the green point sitting at $Re(z) = 0$. The critical point $\sigma_1$ is represented by the closest green point to
$\sigma_0$ to the left. The unstable thimbles are displayed in magenta.

\item The very same reasoning applied to $\frac{\mu}{m} = 0.4$ and $\frac{\mu}{m} = 1.4$ can be applied to
$\frac{\mu}{m} = 0$ and $\frac{\mu}{m} = 1.8$ (actually at $\frac{\mu}{m} = 0$ there is no sign problem and thimble
regularization is not even needed).
\end{enumerate}

We have computed by one-thimble simulations the Taylor coefficients for the scalar condensate
at $\frac{\mu}{m} = 0.4$ and $\frac{\mu}{m} = 1.4$ respectively up to orders $2$ and $5$.
Then we have constructed a (multi-points) Pad\'e approximation using these coefficients as inputs, adding as extra constraints
the coefficients of order $0$ at the boundaries of the region we have considered,
i.e. $\frac{\mu}{m} = 0$ and $\frac{\mu}{m} = 1.8$. These extra constraints were also calculated by one-thimble simulations.

The results are shown in the bottom left picture of fig. \ref{fig:nr5}. The expansion points are displayed as black points.
The numerical results from the Pad\'e approximants are displayed in
black (with error bars): they are in good agreement
with the analytical solution (the black line).
The bottom right picture of fig. \ref{fig:nr5} shows the expansion points (the black points) on the complex $\frac{\mu}{m}$ plane.
The picture also shows the pole of the approximant (the blue point) and the true singularity of the observable (the green point).
The first matches quite well the latter.

\FloatBarrier
%\begin{figure}[tbp]
\FloatBarrier

Finally the simulations have been repeated towards the continuum limit. For this theory the continuum limit 
is reached by increasing $L = \frac{1}{Ta}$ and $\beta = \frac{1}{2g^2a}$ while keeping fixed
the dimensionless products $L \hat{m}$ and $\beta \hat{m}$.
In particular we kept constant $L \hat{m} = 16$ and $\beta \hat{m} = 2$.
The parameters used in the simulations are summarized in tab. \ref{tab:clparams}.

For the finer lattices we proceeded as we did for the $L=8$ lattice. We selected two suitable expansion points
and we calculated the Taylor coefficients up to order $2$ and $5$ by one-thimble simulations. 
As extra constraints we also calculated the coefficients of order $0$ at the boundaries of the $\hat{\mu}$ range.
These data were used as inputs to construct the Pad\'e approximants. For the $L=32,64$ lattices the statistical error
on the $5$th order Taylor coefficient was quite large and this resulted in a large indetermination on the
Pad\'e approximant itself. Fortunately in building our (multi-points) 
Pad\'e approximants we have two different handles
that we can make use of: one handle is the order of the Taylor
expansions, the second one is the number of expansion points. 
For the  $L=32,64$ lattices we decided to use an additional extra constraint in the low $\hat{\mu}$ region in place
of the $5$th order Taylor coefficient.

\begin{figure}[h!]
        \centering
        \includegraphics[width=\textwidth, height=\textheight, keepaspectratio]{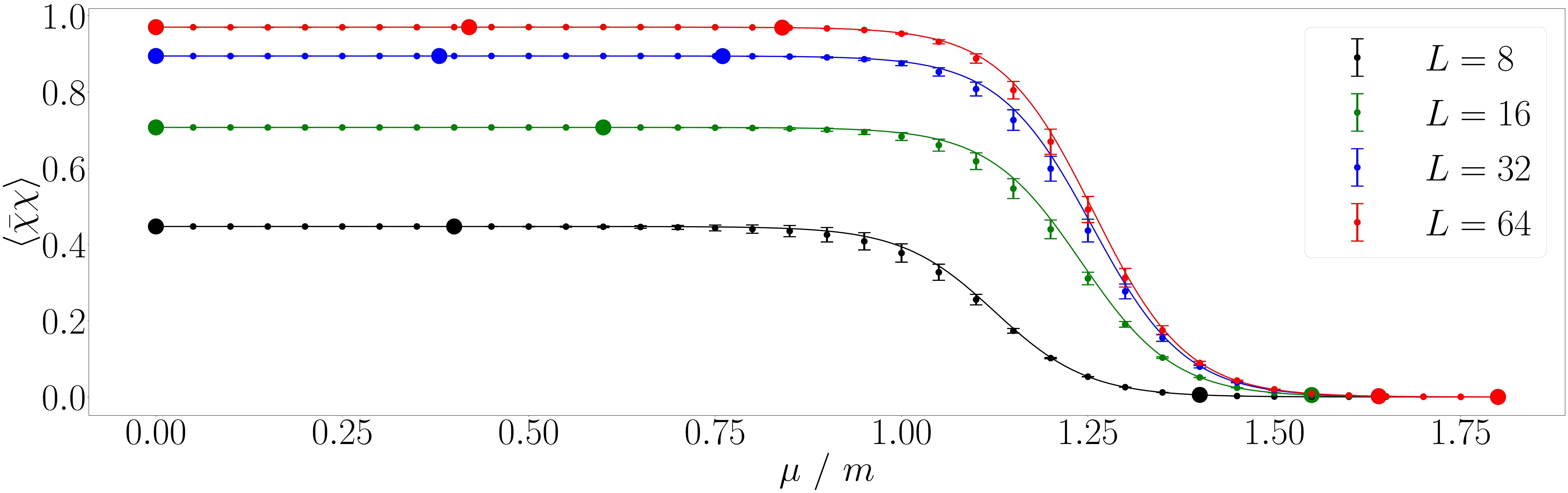}
        \includegraphics[width=\textwidth, height=\textheight, keepaspectratio]{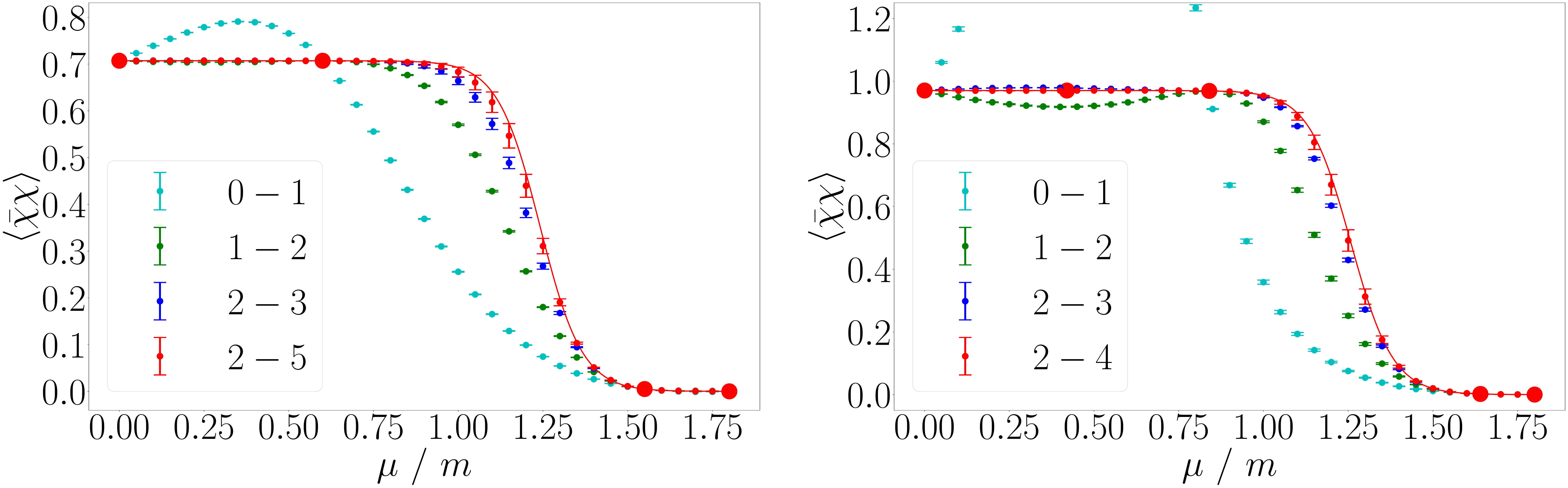}
        \caption{\label{fig:continuum} Results from the continuum limit analysis ($L \hat{m} = 16$, $\beta \hat{m} = 2$). The bottom left and bottom right pictures
                                 show how the Pad\'e approximants converge to the correct solution for the $L=16$ and $L=64$ lattices.
                                 The expansion points are shown as red points. Different colors are used to denote the Pad\'e approximants
                                 obtained by using a gradually increasing order of the Taylor coefficients for the two central expansion points.
                                 Respectively two and three extra constraints on the $0$th order Taylor coefficients were used for the two lattices
                                 and the highest order Taylor coefficient that was calculated is respectively $5$ and $4$. The top picture
                                 summarizes the results we obtained in our analysis. Different colors correspond to different lattice sizes.
                                 The analytical solutions are denoted by the colored solid lines.}
\end{figure}

The numerical results are shown in fig. \ref{fig:continuum}. The
bottom left and bottom right pictures show how the Pad\'e approximants
(the colored error bars) converge to the analytical solution (the red line) as we gradually increase the order of
the Taylor coefficients calculated at the two (central) expansion points. Specifically these pictures illustrate
what happens in the two different cases ($L=16$ and $L=64$) where respectively two and three extra constraints were
used and the highest order that has been calculated in the Taylor expansion was respectively $5$ and $4$. 

The top picture of fig. \ref{fig:continuum} summarizes the results we have obtained for all the lattices.
Different colors denote different lattices. The analytical solutions are drawn as colored solid lines.
The results from the simulations are in good agreement with the analytical solutions
and the statistical error are well under control up to $L=64$.

\begin{table}
\begin{center}
\begin{tabular}{ | c | c | c | }
  \hline
  $L$ & $\beta$ & $ma$ \\ \hline
  $8$ & $1.0$ & $2.00$ \\
  $16$ & $2.0$ & $1.00$ \\
  $32$ & $4.0$ & $0.50$ \\
  $64$ & $8.0$ & $0.25$ \\ \hline
\end{tabular}
\end{center}
\caption{Parameters used for the continuum limit analysis ($L \hat{m}= 16$, $\beta \hat{m} = 2$).}
\label{tab:clparams}
\end{table}

\FloatBarrier
%\begin{figure}[tbp]
\FloatBarrier

\section{Conclusions}
\label{sec:Conclusions}
The Thirring model has been extensively studied as a playground to
test our ability to tackle finite-density theories plagued by a sign problem.
This work aimed at settling an old story: is the thimble approach really
failing for the (supposedly simple) Thirring model? \\
Thimble regularisation of lattice field theories is a conceptually
nice solution to the sign problem, but it can be strongly limited by 
the need for multi-thimbles simulations. The Thirring model has indeed been
a first example of the failure of the simplest application of the
thimble approach, {\em i.e.} the one based on simulations on the
dominant thimble alone. 
While there was little doubt of the conceptual solution to this failure
(other thimbles provide an important contribution), an explicit proof
of this has been till now missing.
We showed that by keeping into account more contributions (on top of
the dominant thimble) the (known) analytical solution can indeed be
reconstructed. While this is conceptually clean, it is from a
numerical point of view quite cumbersome. As a result, the
computations we reported on were restricted to a 
system of modest size. \\
There is nevertheless a more powerful method to achieve the result we
were interested in, which in the end avoids multi-thimbles
simulations. We showed that we could successfully take the continuum
limit in the computation of the Thirring model making use of the
recently introduced method of computing Taylor expansions on
thimbles. After calculating multiple Taylor expansions around points 
where the single-thimble approximation holds true, we could bridge
these different (disjoint) regions by computing (multi-points) 
Pad\'e approximants. The latter also gave us access to the
singularity structure of the theory. While this settles the old story
of the Thirring model in the thimble approach, at the same time
suggests that a similar strategy can be successfully  applied to problems beyond
thimbles. Bridging different
regions by Pad\'e approximants can be an effective way of studying the
phase diagram of a theory for which direct computations can be
approached only in given regions of the parameter space.

\section*{Acknowledgments}
\par\noindent
This work has received funding from the European Union’s Horizon 2020 
research and innovation programme under the Marie Sklodowska-Curie 
grant agreement No. 813942 (EuroPLEx).
We also acknowledge support 
from I.N.F.N. under the research project {\sl i.s. QCDLAT}.
The numerical work for this research has made use of the
resources available to us under the INFN-CINECA agreement. 
It also benefits from the HPC (High Performance Computing) 
facility of the University of Parma, Italy. 

%\newpage

\end{document}